\documentclass[fleqn,usenatbib]{mnras}

\usepackage{newtxtext,newtxmath}
\usepackage[T1]{fontenc}

\DeclareRobustCommand{\VAN}[3]{#2}
\let\VANthebibliography\thebibliography
\def\thebibliography{\DeclareRobustCommand{\VAN}[3]{##3}\VANthebibliography}

\usepackage{graphicx}	% Including figure files
\usepackage{amsmath}	% Advanced maths commands
\usepackage[export]{adjustbox}

\title[Forced eccentricity in a retrograde disc]{Disc breaking through forced eccentricity growth}

\author[M. Overton et al.]{
Madeline Overton$^{1,2}$\thanks{E-mail: madeline.overton@unlv.edu}, Rebecca G. Martin$^{1,2}$, Stephen H. Lubow$^3$, and Stephen Lepp$^{1,2}$
\\
$^{1}$Nevada Center for Astrophysics, University of Nevada, Las Vegas, 4505 South Maryland Parkway, Las Vegas, NV 89154, USA\\
$^{2}$Department of Physics and Astronomy, University of Nevada, Las Vegas 4505 South Maryland Parkway, Las Vegas, NV 89154, USA\\
$^3$Space Telescope Science Institute, 3700 San Martin Drive, Baltimore, MD 21218, USA
}

\date{Accepted XXX. Received YYY; in original form ZZZ}

\pubyear{\the\year{}}

\begin{document}
\label{firstpage}
\pagerange{\pageref{firstpage}--\pageref{lastpage}}
\maketitle

\begin{abstract}
Motivated by misaligned discs observed in eccentric orbit Be/X-ray binaries, we examine the evolution of a retrograde disc around one component of an eccentric binary with hydrodynamic simulations, $n$-body simulations and linear theory. Forced eccentricity growth from the eccentric orbit binary causes the initially circular disk to undergo eccentricity oscillations.  A retrograde disc becomes more radially extended, more highly eccentric and undergoes more rapid apsidal precession compared to a prograde disc.  We find that a retrograde disc can be subject to disc breaking where the disc forms two rings with different eccentricities and longitude of periastrons while remaining coplanar. This could have implications for  the lightcurves  and the X-ray outbursts observed in Be/X-ray binaries.

\end{abstract}


%%%%%%%%%%%%%%%%%%%%%%%%%%%%%%%%%%%%%%%%%%%%%%%%%%

%%%%%%%%%%%%%%%%% BODY OF PAPER %%%%%%%%%%%%%%%%%%

\section{Introduction}

Be/X-ray binaries consist of a high mass Be main sequence star and a compact object companion, which is often a neutron star \citep[e.g.][]{Negueruela2001,Negueruela2001b,Okazaki2002,Reig2011,Okazaki2013,Kretschmar2013_update}. The Be star possesses a Keplerian decretion disc \citep{Lee1991,Pringle1991,Hanuschik1996,Carciofi2011,Rivinius2013} formed, in part, due to its rapid rotation at a high fraction of its critical rotation rate \citep{Slettebak1982,Porter1996,Porter2003}. 
Be stars discs are often observed to be misaligned to the binary orbital plane \citep[e.g.][]{Hummel1998,Hughes1999,Hirata2007,martin2011be}.
A supernova kick  during the formation of the compact object companion can cause the binary orbit to become eccentric and significantly misaligned or even retrograde relative to the Be star decretion disc \citep{Brandt1995, Martinetal2009b,Salvesen2020}.
Additionally, a disc that forms with a sufficiently large relative inclination to the binary orbital plane can undergo nodal precession and align towards retrograde coplanar \citep[e.g.][]{Papaloizou1995,Lubow2000,Bateetal2000}.

Previously, \cite{Overtonetal2023} investigated the dynamics of a retrograde circumprimary disc in a circular orbit binary. The retrograde disc remains quite circular compared to the prograde disc, which is subject to eccentricity growth by the 3:1 resonance because of the large mass ratio binary \citep{Lubow1991,Lubow1992,Murray1998,Goodchild2006,Franchini2019b, Ohanaetal2025}. The lack of ordinary Lindblad resonances in a retrograde disc results in a larger \textit{outer} disc radius that leads to closer interaction with the companion neutron star \citep{Lubowetal2015,Miranda2015}. For a circular orbit binary, \cite{Overtonetal2023} found that the larger disc can be unstable to tilting in the outer regions. In this work, we consider the additional effects of an eccentric binary on the Be star disc dynamics. A test particle around one component of an eccentric binary experiences a forced eccentricity due to the binary potential \citep{Murray&Dermott1999book}. A disc may undergo global oscillations driven by forced eccentricity \citep[e.g.][]{Martin2020alphacen}. Thus, the tilting mechanism described in \cite{Overtonetal2023} may be unable to operate and the retrograde disc behavior is likely to be significantly different between the circular and eccentric binary cases. 

In this Letter we investigate the evolution of a retrograde disc in an eccentric Be/X-ray binary. We show that the large radial extent of a retrograde disc can lead to a coplanar disc undergoing breaking as a result of the forced eccentricity growth. The disc breaks into rings with different eccentricities. While disc breaking has been previously been investigated in detail in misaligned discs as a result of differential nodal precession \citep[e.g.][]{Larwood1997,Nixon2012,Nixon2016,Nealonetal2022}, in this work we demonstrate that disc breaking can occur as a result of eccentricity growth in a coplanar disc.
In Section~\ref{sec:sims} we describe the simulation setup and discuss the disc dynamics for prograde  and retrograde discs  with respect to the eccentric binary orbit. We supplement the hydrodynamical simulation results by investigating the behavior of test particle orbits in Section~\ref{sec:rebound}. We outline our results and draw our conclusions in Section~\ref{sec:conclusion}. 

\section{Hydrodynamical Simulations}
\label{sec:sims} 

In this Section we first describe the smoothed particle hydrodynamics (SPH) simulation setup and then we discuss the results for the prograde and retrograde disc cases. 

\vspace{10mm}

\subsection{Simulation Setup}
\label{sec:setup}

We model coplanar Be/X-ray binaries with retrograde and prograde circumprimary discs using the {\sc phantom} SPH code \citep{Price2010,Lodato2010,Price2018}. 
The Be star and the neutron star companion are represented by sink particles. The Be star has a mass of $M_* = 18 \ M_\odot$, with a sink radius of $r_* = 8 \ R_\odot$ and the neutron star has a mass of $M_{\rm NS} = 1.4 \ M_\odot$ with a sink radius of $r_{\rm NS} = 1 \ R_\odot$. Any particles that move inside the sink radius are accreted, and add their mass and angular momentum to the sink particle \citep{Bateetal1995}. Therefore, we do not model the inner region of the neutron star's accretion disc. 
The binary separation is $a_{\rm b}=95 \,\rm R_\odot$ and the eccentricity is $e_{\rm b} = 0.34$. This gives an orbital period of $P_{\rm orb} = 24.3$ days. This binary setup is modeled after the relatively close, moderate eccentricity Be/X-ray binary 4U 0115+63 with the same orbital parameters ($P_{\rm orb} = 24.3$ days, $e_{\rm b} = 0.34$) \citep{Giacconi1972,Rappaport1978}. We also choose these parameters for comparison to our previous SPH simulations of coplanar, circular orbit Be/X-ray binaries \citep{Overtonetal2023}. The parameters for the binary and the initial circumprimary disc are summarized in Table~\ref{tab:param_table}.

The Be-star disc initially has a total mass of $M_{\rm di}=10 ^{-8} \, \rm M_{\odot}$ with $5 \times 10^5$ particles. The disc is circular and extends from $R=8$ to $30 \, \rm R_{\odot}$. No mass is added to the disc during the simulation, therefore, the inner parts of the disc represent an accretion disc rather than a decretion disc. However, as the disc spreads outwards, the outer parts of the disc behave as a decretion disc \citep[e.g.][]{Pringle1981,MartinandLubow2011}. Since we do not inject particles, the disc loses mass throughout the simulation as it dissipates. Particles from the disc may be accreted onto either star, or be ejected to form circumbinary material \citep[e.g.][]{Franchini2019}. 

The initial surface density profile follows  $\Sigma =\Sigma_0 (r_\star/R)^{3/2}[1-(r_\star/R)^{1/2}]$, where $\Sigma_0$ is a scaling constant that is determined by the initial disc mass. The sound speed is given by a locally isothermal equation of state of the form 
\begin{equation}
    \label{cs}
    c_{\rm s} = c_{\rm s0} \bigg( \frac{M_*}{R} + \frac{M_{\rm NS}}{R_{\rm NS}} \bigg)^{q}
\end{equation}
\citep{Farris2014} with $q=0.5$, where $R$ and $R_{\rm NS}$ are the distances from the primary and secondary objects, respectively. The constant of proportionality $c_{\rm s0}$ is chosen to give the disc aspect ratio, $H/R=0.01$, at the inner disc radius ($R = r_* = 8 \, \rm R_{\odot}$). The temperature profile in the disc is given by $T\propto c_{\rm s}^2$. The sound speed is $c_{\rm s} = H \Omega$ where $\Omega = \sqrt{GM_*/R^3}$ is the Keplarian angular velocity. This gives an approximately constant disc aspect ratio of $H/R \approx 0.01$, since the sound speed near the Be star is $c_{\rm s} \propto R^{-q}$. Thus, the approximate initial mean smoothing length over the disc scale height is $\left<h\right>/H  = 0.9$. A high disc aspect ratio has been shown to weaken the effects of tidal truncation \citep{Martin2024}. In this work we study the interaction of the disc with the companion, so a low disc aspect ratio is preferred. This  also allows for comparison to previous work. We also ran simulations with $H/R = 0.02$, however the disc lifetime was too short to make comparisons, so we do not include the results of those simulations in this work. 

Be star discs are observed to have a large \cite{SS1973} viscosity parameter of $\alpha \approx 0.3$ \citep[e.g.][]{Carciofi2012, Rimulo2018, Ghoreyshi2018}. This is expected for a fully ionised disc \citep{Martin2019}. We use the method outlined in \cite{Lodato2010} to modify the SPH artificial viscosity. Using an artificial viscosity parameter $\alpha_{\rm AV} = 3.32$, the approximate \cite{SS1973} viscosity parameter is given by 
\begin{equation}
    \alpha \approx \frac{1}{10}\alpha_{\rm AV}\frac{\left<h\right>}{H}
\end{equation}
\citep[e.g.][]{Okazaki2002}.

We consider two cases for the Be star disc. The disc is initialized as either prograde with an inclination of $i=0^\circ$, or retrograde with an inclination of $i=180^\circ$. The simulations are allowed to run for a time of $t=75 \, P_{\rm orb}$.

In order to analyse properties of the disc in our simulations, we bin the particles into 100 bins in particle semi-major axis evenly distributed between $a=8 \, R_{\odot}$ and $a = 100 \, R_{\odot}$. We choose to bin by semi-major axis rather than instantaneous radius because the disc becomes very eccentric. The disc properties at each semi-major axis are calculated by averaging over the particle properties in each bin.

\begin{table}
	\centering
	\caption{Parameters for the binary and the circumprimary disc.}
	\label{tab:param_table}
	\begin{tabular}{lccr} 
		\hline
		  Binary parameters & Symbol/Units & Value \\
		\hline
		  Mass of Be Star & $M_*/ M_{\odot}$ & 18  \\
            Sink Radius of Be Star & $r_* / R_{\odot}$ & 8 \\
		  Mass of NS companion & $M_{\rm NS}/ M_{\odot}$ & 1.4 \\
            Sink radius of NS & $r_{\rm NS} / R_{\odot}$ & 1 \\
            Orbital period & $P_{\rm orb}/ \rm day$ & 24.3 \\
            Semi-major axis & $a_{\rm b}/R_{\odot}$ & 95 \\
            Eccentricity & $e_{\rm b}$ &  0.34 \\
            \hline 
            Initial Circumprimary Disc Parameters & Symbol & Value \\
            \hline 
		  Disc mass  & $M_{\rm di}/M_{\odot}$ & $10^{-8}$ \\
            Inclination to the binary orbital plane & $i$ & $0$, $180^{\circ}$ \\
            Initial disc outer radius & $R/R_{\odot}$ & 30 \\
            Shakura \& Sunyaev viscosity parameter & $\alpha$ & 0.25-0.39 \\
            Aspect ratio & $H/R \, (R = r_*)$ & 0.01 \\
		\hline
	\end{tabular}
\end{table}

\begin{figure*}
    \centering
    \includegraphics[width=0.75\textwidth]{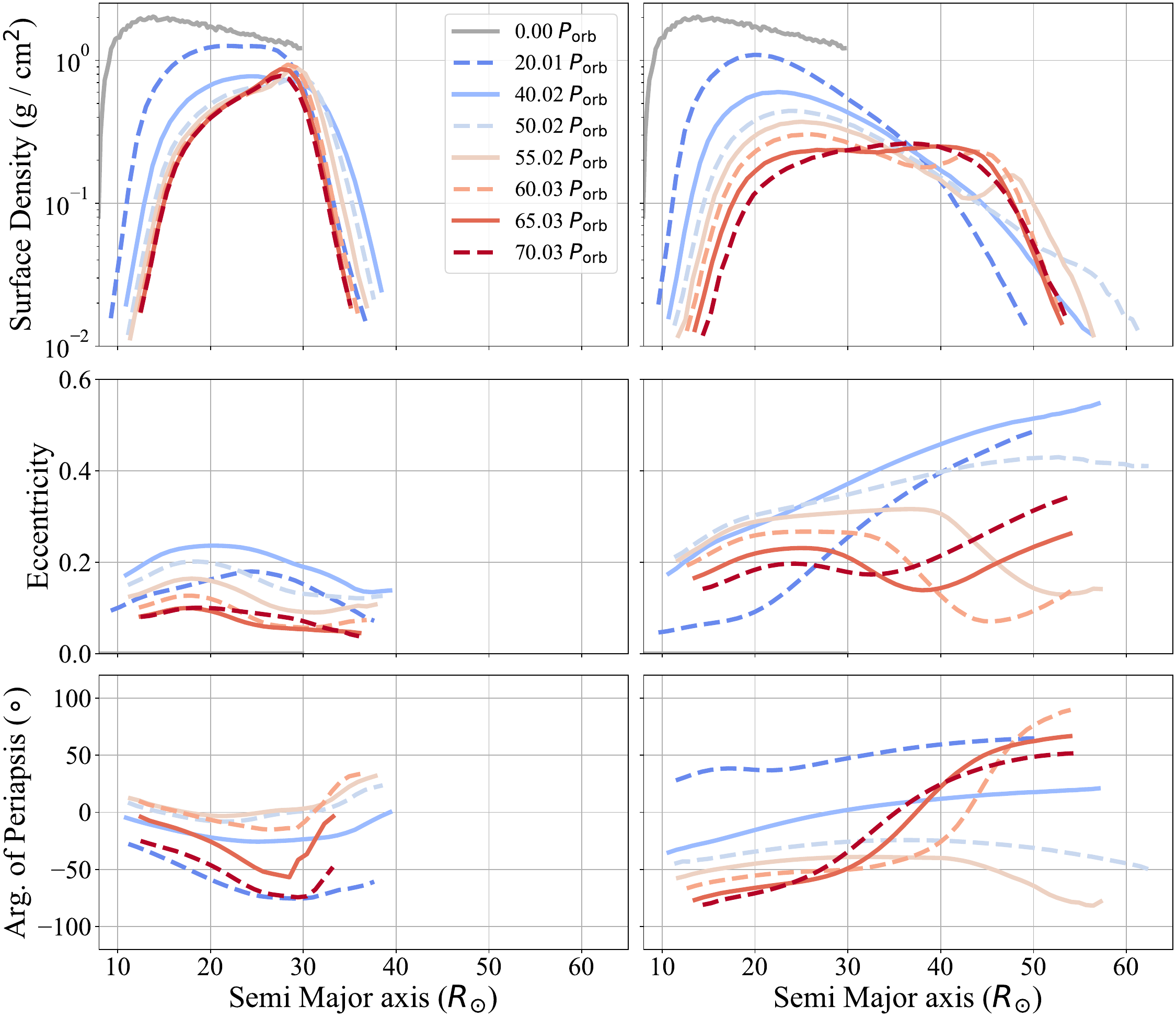}
       %\vspace{1cm}
    \caption{Surface density of the Be star disc (top panels), eccentricity of the disc (middle panels), and the argument of periapsis (bottom panels) as a function of semi major axis, with different lines representing different times in the simulation. The prograde disc results are shown on the left panels and the retrograde on right side panels. We plot the data for portions of the disc with a surface density greater than approximately $10^{-2} \, \rm g \ cm^{-2}$. }

    \label{fig:surf_rho}
\end{figure*}

\begin{figure*}
    \centering
    \includegraphics[width=0.8\textwidth]{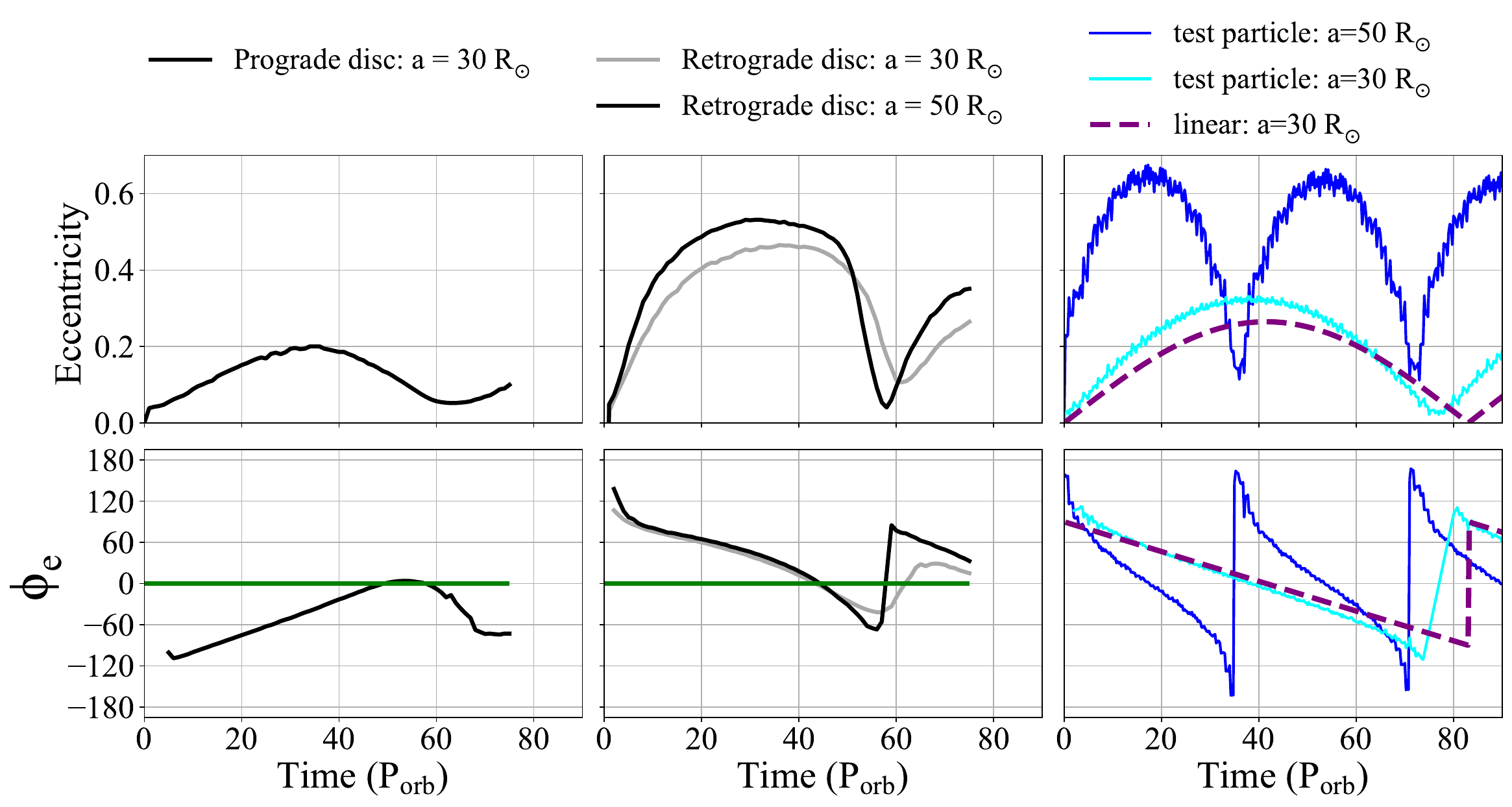} 
    \caption{The eccentricity (upper row) and argument of periapsis (lower row) for the prograde disc (left column), the retrograde disc (middle column), and the  test particles (right column). The argument of periapsis of the binary is constant at $\phi_e = 0$, represented by the green line in the lower left and middle columns. The argument of periapsis is not well defined for a circular orbit, so we only plot $\phi_e$ for the disc and test particles when $e>0.05$. }
    \label{fig:eephi}
\end{figure*}

\subsection{Prograde disc}
\label{sec:prograde}

The evolution of a prograde disc around one component of an eccentric binary has been thoroughly investigated in previous work \citep[e.g.][]{Lubow1991, Lubow1991b, Okazaki2001, Okazaki2002,Martinetal2014, Franchini2019b,Martin2020alphacen}. We provide this case here only for direct comparison to the retrograde disc (section \ref{sec:retrograde}). 

The left panels of Fig.~\ref{fig:surf_rho} show the evolution of the surface density, eccentricity, and the argument of periapsis (from top to bottom) for the prograde disc as a function of particle semi-major axis. We plot the disc properties for several times, focusing on the later part of the simulation. 
We plot the data for portions of the disc with a surface density greater than approximately $10^{-2} \, \rm g \ cm^{-2}$. The argument of periapsis is defined with
\begin{equation}
	\label{phi_e}
	\phi_e = \arctan \bigg( \frac{e_{y}}{e_{x}}\bigg)
\end{equation}
for coplanar orbits, where $e_x$ and $e_y$ are the $x$ and $y$ components of the eccentricity vector
\begin{equation}
% Trimane planetary dynamics pg 9
\boldsymbol{e}= \frac{\big(\boldsymbol{\dot{r}} \times (\boldsymbol{r} \times \boldsymbol{\dot{r}}) \big)}{ GM_*} - \frac{\boldsymbol{r}}{r},
\end{equation}
projected into the $x-y$ plane (the binary orbital plane), where $\boldsymbol{r}$ is the position vector of a particle, $G$ is the gravitational constant, and $M_*$ represents the mass of the primary star, which hosts the disc.
The argument of periapsis is not well defined for a circular disc, so we only plot $\phi_e$ for parts of the disc where the eccentricity is greater than $0.05$.

As shown in the top left panel of Fig.~\ref{fig:surf_rho}, the disc relaxes from its initial outer radius and surface density as it dissipates. The disc expansion timescale corresponds to the viscous timescale. However, the disc is prevented from significantly expanding due to its interaction with the tidal potential \citep[e.g][]{Artymowicz1994, Okazaki2002, Miranda2015}.
A disc is said to be truncated where the surface density experiences a sharp drop off. We find that the prograde disc is tidally truncated at a semi major axis of approximately $30 \, R_{\odot}$.

The disc is initially circular but experiences forced eccentricity growth due to the interaction with the tidal potential of the eccentric binary. As shown in the middle left panel of Fig.~\ref{fig:surf_rho}, the strength of the eccentricity forcing varies as a function of disc radius \citep[see also][]{AndradeInesetal2016, Quarlesetal2018}. 
The lower left panel of Fig.~\ref{fig:surf_rho} demonstrates the outer regions of the prograde disc experience faster precession compared to the innermost regions.

Fig.~\ref{fig:eephi} plots the eccentricity (upper row) and argument of periapsis (lower row) as a function of time, at a semi major axis of $a=30 \, R_{\odot}$ for the prograde disc (left column). The argument of periapsis of the binary is also shown on the plot represented by the solid green line, which stays constant at $\phi_e = 0^\circ$. The eccentricity oscillates in time reaching a peak at a time of approximately $t=35 \, P_{\rm orb}$. A change in the argument of periapsis indicates apsidal precession. In the prograde disc case we see prograde apsidal precession.

\subsection{Retrograde disc}
\label{sec:retrograde}

The retrograde circumprimary disc experiences quite different behavior in response to the tidal potential of the binary, when compared to the prograde case. 
The right  panels of Fig.~\ref{fig:surf_rho} show the evolution of the surface density, eccentricity, and the argument of periapsis (from top to bottom) for the retrograde disc as a function of particle semi-major axis at various times throughout the simulation. As seen the top panels of Fig. ~\ref{fig:surf_rho},  
the retrograde circumprimary disc extends much closer to the binary orbit ($a_{\rm b} = 95 \, R_{\odot}$) than the prograde case. The disc immediately begins to expand from its initial radius of $30 \, R_{\odot}$ and dissipates throughout the entirety of the simulation. Since the initial disc radius is much smaller than where the disc is truncated, the disc must undergo a longer period of expansion before being truncated. 
By a time of $t=50 \, P_{\rm orb}$ the outer portions of the disc experience a sharp drop off in the surface density, signifying the disc is truncated. Between times $t=50 \, P_{\rm orb}$ and approximately $t=60 \, P_{\rm orb}$, the disc’s truncation radius shrinks from $a=55 R_{\odot}$to $a=45 \, R_{\odot}$, at which point it is also more strongly truncated. The retrograde disc is larger than the prograde case (section \ref{sec:prograde}) which is truncated at only $a \approx 30 \, R_{\odot}$.

The truncation radius may become smaller due to the orientation of the disc eccentricity with respect to the binary orbit eccentricity. 
The disc experiences retrograde apsidal precession with respect to the binary orbit. Note that this is opposite the direction of the precession relative to the binary in the prograde case. 
In Fig.~\ref{fig:eephi} we plot the eccentricity (upper row) and argument of periapsis $\phi_{e}$
(lower row) as a function of time, at a semi major axis of $a=30$ and $50 \, R_{\odot}$ for the retrograde circumprimary disc (middle column).
As shown in the bottom middle panel of Fig.~\ref{fig:eephi}, while the truncation radius is shrinking (between $t=50$ and $60 \, P_{\rm orb}$) the argument of periapsis of the disc is moving further out of alignment with the binary’s argument of periapsis (shown by the solid green line) which stays constant at $\phi_e = 0^{\circ}$. Thus, the companion has a closer approach to the outer edge of the disc, therefore restricting the size of the disc.

As seen in the bottom right panel of Fig.~\ref{fig:surf_rho}, the argument of periapsis rapidly changes after $t=55 \, P_{\rm orb}$. Prior to $55 \, P_{\rm orb}$, $\phi_{e}$ is relatively constant as a function of radius indicating the entire disc precesses together. However, the disc undergoes a breaking event and afterwards, the disc develops an inner and outer disc where the argument of periapsis is separated by more than $90^{\circ}$. The change in $\phi_{e}$ also corresponds to a change in the disc’s eccentricity profile (shown in the middle right panel of \ref{fig:surf_rho}). For much of the simulation, the outer regions of the disc are significantly more eccentric than the inner regions. However, by $55 \, P_{\rm orb}$ the disc eccentricity in the outer regions dips relatively low to $e \approx 0.1$ at a semi-major axis of $a=50 \, R_{\odot}$. 
Fig.~\ref{fig:eephi} emphasizes that the circularization of the disc coincides with the jump in $\phi_{e}$. The disc becomes highly eccentric, with a peak of $e \approx 0.5$ (at semi-major axis of $a = 50 \ R_{\odot}$), however at approximately $t=55\,P_{\rm orb}$, the disc suddenly becomes nearly circular, before the eccentricity begins to grow again. If there were enough time, and the disc did not dissipate away entirely, it appears that the eccentricity could continue to oscillate. We explain the oscillations through comparison to test particle orbits in Section~\ref{sec:rebound}.

\begin{figure*}
    \centering
    \hspace{1 in} \includegraphics[width=0.85\textwidth]{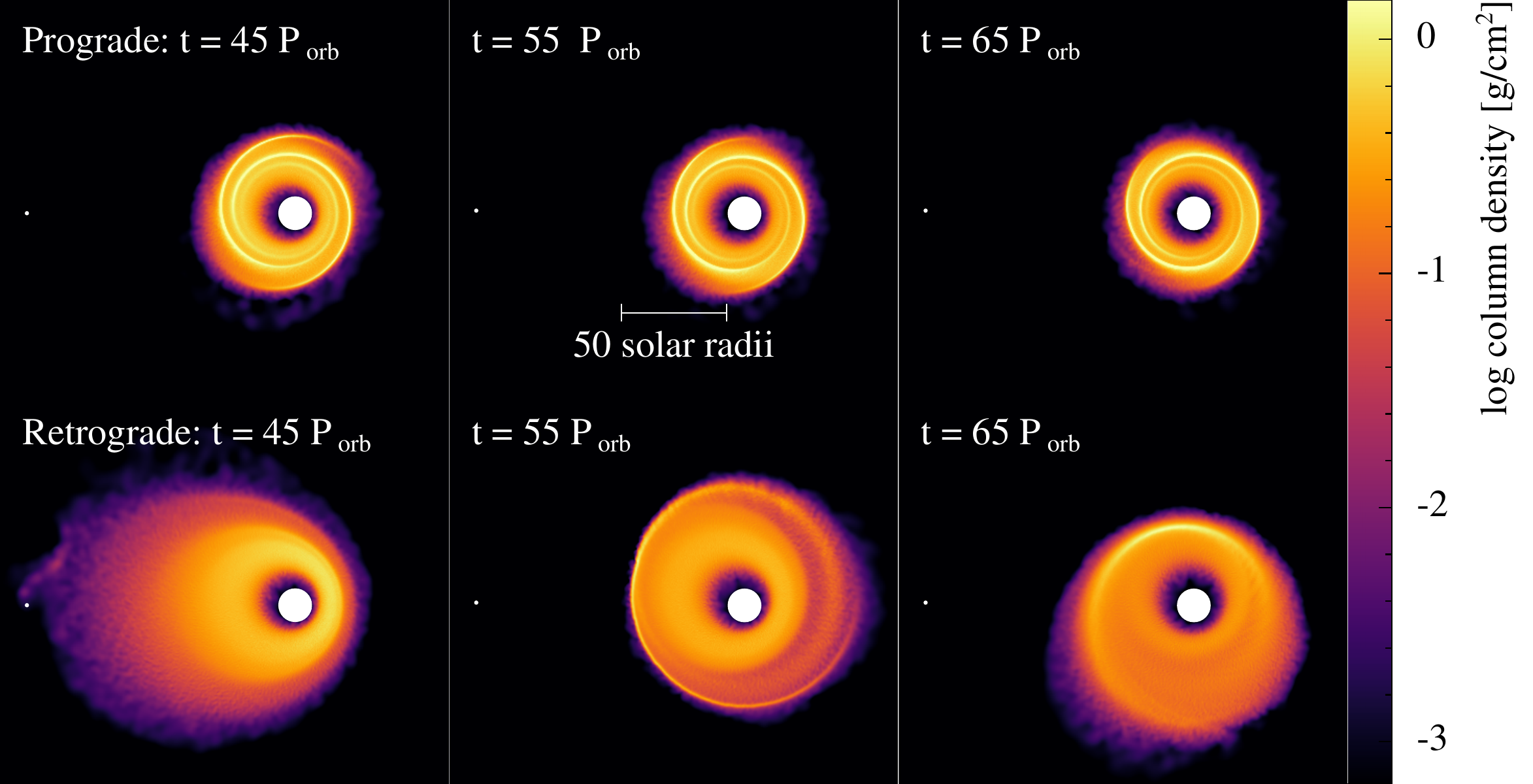}
    \includegraphics[width=0.84\textwidth]{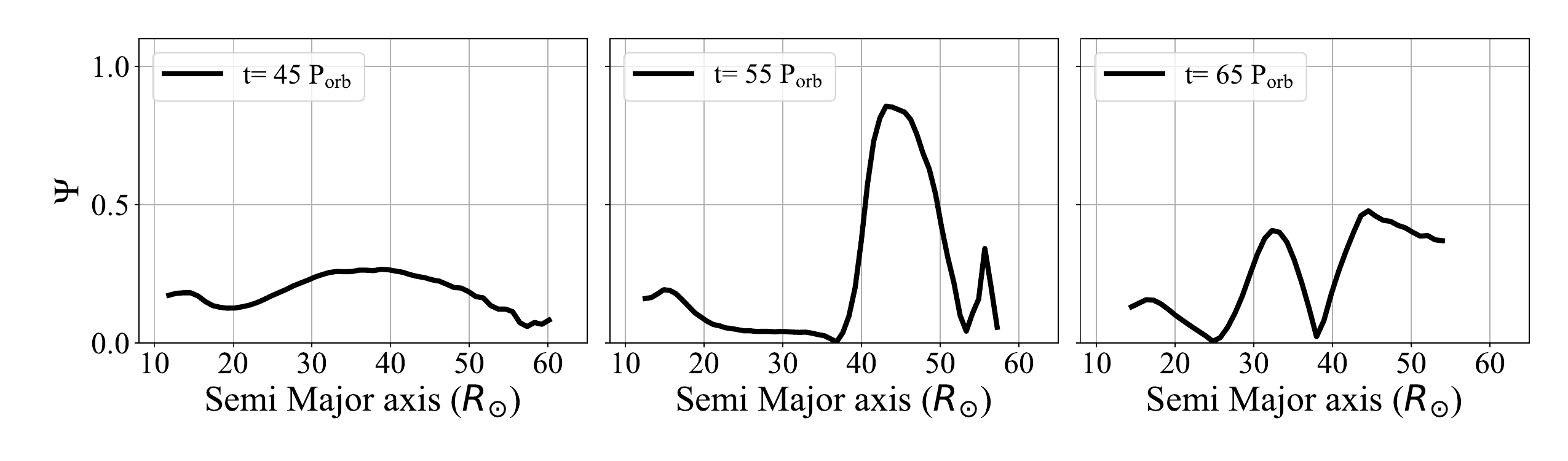} \hspace{0.05 in}
    \caption{The column density of the prograde Be star disc (upper row) and retrograde Be star disc (middle row) at times $t= 45$ (left), 55 (middle) and $65 \, P_{\rm orb}$ (right) in the $x-y$ plane in which both the Be star and the neutron star orbit. The bottom row shows the dimensionless quantity $\Psi$ at the same times for the retrograde disc. Only the values of $\Psi$ are shown where the surface density is greater than $10^{-2}$ g cm$^2$.}
    \label{fig:splash3}
\end{figure*}

During the abrupt change of the outer disc eccentricity, the disc appears to undergo a break. The top right panel of Fig.~\ref{fig:surf_rho} shows the dip in the surface density near $a=45 \, R_{\odot}$, at a time $t = 55 \, P_{\rm orb}$. For an inclined disc, a commonly used break criteria is a maximum in the warp profile \citep[e.g.][]{Lodato2010, Nixon2012, Nixon2015, Nealonetal2022, Doganetal2023}. For this work considering coplanar discs, we instead consider the breaking criteria where there is a local minimum in the surface density and a maximum in the dimensionless quantity 
\begin{equation}
    \label{azimuthal_warp}
    \Psi =  a\bigg| \frac{\partial e}{\partial a} \bigg|
\end{equation} 
where $e = |\boldsymbol{e}|$. The quantity $\Psi$ for a coplanar disc is analogous to the warp profile for a tilted disc. A large value of $\Psi$ represents a sudden jump in the eccentricity. The retrograde disc possesses a local minimum in the surface density and a local maximum in $\Psi$, at the same location in the disc, between times of
$t = 53 \, P_{\rm orb}$ and $t = 66 \, P_{\rm orb}$. However, by a time of $t=65\,P_{\rm orb}$, the inner disc also circularises and radial communication is restored so that the disc properties recover a smooth profile. The quantity $\Psi$ is plotted in the bottom row of Fig.~\ref{fig:splash3} for times $t=45$, $55$, and $65 \, P_{\rm orb}$. At $t=45 \, P_{\rm orb}$ there is no break and therefore $\Psi$ is relatively low and constant throughout the disc. At $t=55 \, P_{\rm orb}$ the disk is near its most broken, and $\Psi$ has a maximum at semi-major axis $a = 43 \, R_{\odot}$. This is in agreement with the surface density profile shown in the upper right panel of Fig.~\ref{fig:surf_rho}. At $t=65 \, P_{\rm orb}$ the break begins to close and  $\Psi$ has decreased. 

Neither the prograde nor retrograde circumprimary disc becomes significantly inclined. \cite{Overtonetal2023} found a tilt instability for a retrograde circumprimary disc in a circular binary occurring in the regions of the disc  $R\gtrsim 50 \, R_{\odot}$, and suggested that the tilt instability is due to a retrograde Lindblad resonance \citep[see also][]{Henon1973}. In this work, the eccentric binary forces the disc to become eccentric, thereby restricting the disc to a smaller radius. Since the eccentric binary with a retrograde disc case is truncated at approximately $a=50 \, R_{\odot}$ it does not reach the location of the resonance and does not experience the tilt instability. 

\section{Test particle orbits }
\label{sec:rebound}

In this Section we consider the evolution of test particles orbits in order to better understand the disc simulations. We first consider $n$-body simulations and then we compare this with linear theory.

\subsection{$n$-body simulations}

We supplement our investigation of the retrograde disc with test particle simulations using the $n$-body integrator, {\sc Rebound} \citep{rebound,reboundias15}. The initial setup is an eccentric binary with the same parameters as described by Table~\ref{tab:param_table} with massless test particles at varying initial separations from the Be star to represent the behavior of thin annuli of the decretion disc. The test particles are initially in circular, Keplerian orbits, which are coplanar and retrograde with respect to the binary.

The right panels in Fig.~\ref{fig:eephi} show the eccentricity (upper panel) and the argument of periapsis (bottom panel) as a function of time for the test particles. The test particles undergo eccentricity oscillations and a jump in the argument of periapsis when the eccentricity is at the minimum of the oscillation. Test particles at larger initial separations have larger eccentricity growth, shorter timescales for eccentricity oscillations, and shorter timescales for apsidial precession. A test particle with an initial separation of $a= 50 \, R_{\odot}$ oscillates between an eccentricity $e = 0.15$ and $e = 0.65$. The test particle at a smaller initial separation ($a=30 \, R_{\odot}$) has smaller magnitude oscillations which reach a lower minimum, with eccentricity between approximately $e = 0.05$ and $e = 0.35$. 

\subsection{Linear theory}

The linear evolutionary equation for the complex eccentricity, $E(\lambda, \, t) = e\, \text{exp}(i \omega(\lambda, \, t))$ is given by 
\begin{equation}
    \begin{aligned}
        \label{evo_e}
        \Sigma (GM_{*}\lambda )^{1/2} \frac{\partial E}{\partial t} = &\frac{\partial}{\partial \lambda} \bigg( Z_1 \Sigma c^2_s \lambda \bigg) + Z_2 \Sigma c^2_s  \\
        &+ \frac{1}{4} \frac{i G M_* \Sigma \beta }{ \lambda_b} \bigg[ b^{(1)}_{3/2}E - b^{(2)}_{3/2}E_b \bigg]
    \end{aligned}
\end{equation}
\citep{Ogilvie2001, Okazaki2002} for an isothermal disc under the influence of a companion with low eccentricity. The semi-latus rectum in plane polar coordinates ($R$, $\phi$) is $\lambda = R (1 + e\text{cos}(\phi - \omega))$ where $\omega = \omega(\lambda, t)$ is the longitude of periastron with respect to the stellar periastron. 
The constants $E_b$ and $\lambda_b$ are the complex eccentricity and semi-latus rectum of the binary, $b_{\gamma}^{(m)} = b^{(m)}_{\gamma}(\beta)$ is the Laplace coefficient of celestial mechanics, and ${\beta = \lambda / \lambda_b}$. 

We consider a retrograde test particle and therefore, the right hand side of equation~(\ref{evo_e}) becomes negative and the dimensionless coefficients $Z_1 = Z_2 = 0$ (since there is no shear or bulk viscosity). 
The linear evolutionary equation arises from neglecting terms of relative order $\mathcal{O}(e^2)$ in the non-linear theory of the evolution of a 3D eccentric accretion disc. For small $e$, $\lambda \approx R$. 
With these modifications, equation~(\ref{evo_e}) becomes 
\begin{equation}
    \label{evo_e2}
    \tau \frac{\partial E}{\partial t} = - iR \bigg[ b^{(1)}_{3/2}E - b^{(2)}_{3/2}E_b \bigg],
\end{equation}
where $\tau = \frac{ 4 R_b \sqrt{GM_*R} }{ G M_{\rm NS}  }$. 
We plot the solution for a test particle with an initial separation $a = 30 \, R_{\odot}$ in the right panels of Fig.~\ref{fig:eephi} with a purple dashed line. Note the similarity to the $n$-body simulation for initial separation $a = 30 \, R_{\odot}$. This indicates that the retrograde disc experiences eccentricity growth due to forced eccentricity from interactions with the tidal potential. The differences between the results of the $n$-body simulation and the linear evolutionary equation may occur when the particle eccentricity becomes too high. At larger initial separations the particle behavior is significantly altered by close interactions with the companion, and can not be fully described by the linear theory. 

As discussed in Section~\ref{sec:retrograde}, the eccentricity of the retrograde disc behaves similarly to the test particles. The disc experiences global oscillations due to forced eccentricity. Also, like the test particles, the outer regions of the disc reach a higher eccentricity and there occurs sudden jumps in the argument of periapsis. However, since the disc experiences viscous communication and other collective effects, the individual annuli of the disc cannot behave independently. Thus, the eccentricity growth in the disc is not as extreme as the test particle with equivalent separation, and the disc can not return to exactly circular, although it does become quite low. 

\section{Conclusions}
\label{sec:conclusion} 

We have compared the evolution of coplanar prograde and retrograde decretion discs in  eccentric Be/X-ray binaries with  hydrodynamical simulations. While the prograde disc quickly reaches a steady state and is truncated near $30 \, R_{\odot}$, the retrograde disc becomes more radially extended. This allows for the retrograde disc to interact more strongly with the companion. This manifests in large forced eccentricity growth primarily at the outer edge of the retrograde disc. As the eccentricity oscillates, the rapid change in eccentricity at the outer edge in conjunction with a change in the argument of periapsis allows a break in the disc to occur. The break occurs between regions of different eccentricity. The break persists for over $10$ orbital periods. This breaking is unique to the more commonly studied breaking scenario, where the disc is inclined out of the binary orbital plane and communication in the disc is broken between mutually inclined regions.

We compared the disc behavior to the orbits of $n$-body test particles and the solution of the linear evolutionary equation for a test particle to confirm the behavior of the disc is due to forced eccentricity from the binary. The retrograde circumprimary disc results in satisfactory conditions for a break due to the radially extended disc and large change in eccentricity. These conditions can not occur for the coplanar prograde case for the parameters of our work because the prograde disc has a smaller truncation radius and eccentricity forcing from the binary does not result in as steep of an eccentricity gradient, when compared to the retrograde case.
This breaking mechanism may be able to occur in other accretion discs given the right conditions: a strong radial change in the eccentricity, and a slow communication timescale over the radial extent of the disc. Recently, with magnetohydrodynamic simulations \cite{Ohanaetal2025} has shown that the prograde circumprimary disc can break into regions of different eccentricity because MRI turbulence amplifies eccentricity in the inner regions of the disc (found earlier by \citep{Chan2022}), while the 3:1 resonance amplifies eccentricity outside this region.

Be stars may continuously eject material into the  disc \citep[e.g.][]{Okazaki2002}. However, in our simulation we begin with a disc and we do not add material to it. While it is beyond the scope of this work, with injection into the simulation, the eccentricity oscillations may continue and the breaking behaviour may repeat. We note that repeated tearing by nodal precession has been seen before in simulations that include particle injection \citep{Suffak2022}. With injection into a retrograde coplanar disc, the inner parts of the disc may be held more circular and the disc could develop an even larger eccentricity gradient. Furthermore, the addition of particles to the disc will improve the resolution. In SPH simulations of misaligned accretion discs \citep{Nealon2015} found that disc breaking is only observed in simulations with sufficiently high resolution and that disc breaking becomes stronger for the highest resolution simulations.

Although we do not provide simulated observables, we expect that extreme variability in the size and eccentricity of the retrograde disc would cause variability in the portion of the emission spectra where there is a contribution from the disc. If the eccentric break is a large enough feature, it may also impact the emission spectra. These effects may be seen in the equivalent width evolution of the H$\alpha$ line \citep[e.g][]{Huang1972, Monageng2017, Suffak2020,Rast2024}.  
Additionally, the rapid changes in the outer edge of the disc will impact the accretion onto the neutron star, which may lead to variability in the X-ray outbursts. If the eccentricity of the disc continued to oscillate, there may be periodicity to the X-ray variability.

\section*{Acknowledgements}
%The Acknowledgements section is not numbered. Here you can thank helpful colleagues, acknowledge funding agencies, telescopes and facilities used etc. Try to keep it short.
We thank Daniel Price for providing the phantom code for SPH simulations and acknowledge the use of SPLASH \citep{Price2007} for the rendering of the figures. The test particle simulations made use of the REBOUND code which is free to download at http://github.com/hannorein/rebound. Computer support was provided by UNLV’s National Supercomputing Center. We acknowledge support from NASA through grants 80NSSC21K0395 and 80NSSC19K0443 and 80NSSC23M0104.

%%%%%%%%%%%%%%%%%%%%%%%%%%%%%%%%%%%%%%%%%%%%%%%%%%
\section*{Data Availability}
%The inclusion of a Data Availability Statement is a requirement for articles published in MNRAS. Data Availability Statements provide a standardised format for readers to understand the availability of data underlying the research results described in the article. The statement may refer to original data generated in the course of the study or to third-party data analysed in the article. The statement should describe and provide means of access, where possible, by linking to the data or providing the required accession numbers for the relevant databases or DOIs.
The data underlying this letter will be shared on reasonable request to the corresponding author.



\bibliographystyle{mnras}
\bibliography{long_cite}

%\begin{thebibliography}{99}
%\end{thebibliography}

%%%%%%%%%%%%%%%%%%%%%%%%%%%%%%%%%%%%%%%%%%%%%%%%%%

%%%%%%%%%%%%%%%%% APPENDICES %%%%%%%%%%%%%%%%%%%%%

% \appendix

% \section{Some extra material}

% If you want to present additional material which would interrupt the flow of the main paper,
% it can be placed in an Appendix which appears after the list of references.

%%%%%%%%%%%%%%%%%%%%%%%%%%%%%%%%%%%%%%%%%%%%%%%%%%

% Don't change these lines
\bsp	% typesetting comment
\label{lastpage}
\end{document}